%% file: main.tex
\documentclass[sigconf,screen]{acmart}

\setcopyright{cc}
\setcctype{by}
\acmDOI{10.1145/3832783.3837489}
\acmYear{2026}
\copyrightyear{2026}
\acmISBN{979-8-4007-2882-2/2026/10}
\acmConference[ASE '26]{Proceedings of the 41st IEEE/ACM International Conference on Automated Software Engineering}{October 12--16, 2026}{Munich, Germany}
\acmBooktitle{Proceedings of the 41st IEEE/ACM International Conference on Automated Software Engineering (ASE '26), October 12--16, 2026, Munich, Germany}
\acmSubmissionID{ase26main-p1306-p}
\received{2026-03-26}
\received[accepted]{2026-06-18}

\input{macro}

\title[Doc2Feat-Bench: Evaluating Documentation-Driven Feature Addition]{Doc2Feat-Bench: Evaluating Documentation-Driven Feature Addition}

\author{Zhonghao Jiang}
\email{zhonghao.j@zju.edu.cn}
\affiliation{
\institution{College of Computer Science and Technology and the State Key Laboratory of Blockchain and Data Security, Zhejiang University}
\city{Hangzhou}
\country{China}}

\author{Le Deng}
\email{dengle@zju.edu.cn}
\affiliation{%
\institution{College of Computer Science and Technology and the State Key Laboratory of Blockchain and Data Security, Zhejiang University}
\city{Hangzhou}
\country{China}}

\author{Jialun Cao}
\email{jialuncao@ust.hk}
\affiliation{%
\institution{Hong Kong University of Science and Technology}
\city{Hong Kong}
\country{China}}

\author{Michael Pradel}
\email{michael@binaervarianz.de}
\affiliation{%
  \institution{CISPA Helmholtz Center for Information Security}
  \city{Stuttgart}
  \country{Germany}}

\author{Zhongxin Liu}
\authornote{Corresponding Author.}
\email{liu\_zx@zju.edu.cn}
\affiliation{%
  \institution{College of Computer Science and Technology and the State Key Laboratory of Blockchain and Data Security, Zhejiang University}
  \city{Hangzhou}
\country{China}}

\ccsdesc[500]{Software and its engineering~Automatic programming}
\keywords{Agents, Documentation-Driven, Feature Addition}

\renewcommand{\shortauthors}{Zhonghao Jiang, Le Deng, Jialun Cao, Michael Pradel, Zhongxin Liu}

\input{Sections/00_abstract}

\begin{document}
\maketitle

\input{Sections/01_introduction}

\input{Sections/02_featurebench}
\input{Sections/03_benchmark_stat}
\input{Sections/04_setting}

\input{Sections/05_results}
\input{Sections/06_discussion}
\input{Sections/07_related_work}

\input{Sections/08_conclusion}

\section*{Data Availability}
The replication package and online Appendix are available at \url{https://doi.org/10.5281/zenodo.21762257}.
Our official repository is available at \url{https://github.com/Doc2Feat-bench/Doc2Feat-bench}.
The leaderboard of \nc is available at \url{http://doc2featbench.org}.
An earlier version of \nc was referred to as NoCode-bench.

\begin{acks}
This research is supported by the National Natural Science Foundation of China (No.92582107), Zhejiang Provincial Natural Science Foundation of China (No.LZ25F020003), and by the German Research Foundation (DFG; projects 492507603, 516334526, and 526259073). 
\end{acks}

\bibliographystyle{ACM-Reference-Format}
\balance
\bibliography{main}

\end{document}

%% file: macro.tex
\usepackage{array}
\usepackage{textcomp}
\usepackage{stfloats}
\usepackage{verbatim}
\usepackage{graphicx} %
\usepackage{pifont}
\usepackage[ruled,vlined]{algorithm2e}
\usepackage{enumitem}
\usepackage{booktabs}
\usepackage{makecell}
\usepackage{todonotes}
\usepackage{url}
\usepackage{listings}
\usepackage{color}
\usepackage{tcolorbox}
\usepackage{multirow}
\usepackage{amsmath}
\usepackage{soul}
\usepackage{ragged2e}
\usepackage{threeparttable}
\usepackage{balance}
\usepackage{xcolor}
\usepackage{colortbl}
\usepackage{tabularx}
\usepackage{adjustbox}
\usepackage{array}
\usepackage{makecell}
\usepackage{adjustbox}
\newcolumntype{C}[1]{>{\centering\arraybackslash}m{#1}}

\definecolor{lo}{HTML}{FAE3D6} 
\definecolor{lb}{HTML}{D9F2D1} 
\usepackage{mdframed}
\newcounter{rq} 
\newcommand{\answerRQ}[1]{\refstepcounter{rq}
\vspace{1mm}
\begin{mdframed}[linecolor=gray,roundcorner=12pt,backgroundcolor=gray!15,linewidth=3pt,innerleftmargin=2pt, 
leftmargin=0cm, rightmargin=0cm, topline=false, bottomline=false, rightline=false]
\textbf{RQ\arabic{rq} Summary:} #1
\end{mdframed}
}
\newcommand{\nc}{{Doc2Feat-Bench}\xspace}
\newcommand{\swe}{{SWE-bench}\xspace}

\newcommand{\paradigm}{documentation-driven feature addition\xspace}

\usepackage{microtype}
\newif\ifrevision
\revisionfalse

\long\def\rev#1{\ifrevision{\color{blue}#1}\else#1\fi}

\newcommand*{\new}[1]{\rev{#1}}

\newcommand{\appendixurl}{\href{https://doi.org/10.5281/zenodo.21762257}{Appendix}}

%% file: Sections/00_abstract.tex
\begin{abstract}
Documentation changes in mature software projects often describe newly introduced or modified behavior.
This makes them a natural basis for \emph{documentation-driven feature addition}, where software engineering agents implement features from public-facing documentation updates.
However, existing benchmarks for agentic software development primarily rely on issue reports written for maintainer coordination and often contain implementation-level details.
Motivated by this observation, this work introduces \nc, a benchmark for documentation-driven feature addition.
\nc consists of 634 tasks across 10 mature open-source software projects, involving about 114k code changes in total.
Each task pairs a documentation change with the corresponding implementation and developer-written tests, enabling evaluation of whether a software engineering agent can implement the feature described by the documentation diff.
The benchmark is constructed through a five-phase pipeline, starting from release notes, and offers broad coverage across libraries, developer tools, and frameworks while preserving real-world development settings.
To facilitate lightweight and reliable evaluation under limited resources, we further curate a human-validated subset named \nc Verified.
It contains 114 high-quality instances whose task clarity and evaluation validity are manually verified.
We use \nc to assess a range of state-of-the-art software engineering agents with two widely used scaffolds.
Experimental results show that despite significant token consumption, the best task success rate remains as low as 37.72\% (OpenHands with Qwen3-Coder-480B).
Our analysis reveals that software engineering agents face key challenges in performing cross-file edits, understanding existing code modules, and correctly interpreting documentation semantics.
\end{abstract}

%% file: Sections/01_introduction.tex
\input{Figures/fig_task}

\section{INTRODUCTION}
Feature addition refers to the activity of introducing new functionality or modifying existing behavior in software systems to meet evolving user needs~\cite{sommerville_software_engineering,hindle2008large}.
It is a central activity in software evolution, accounting for a large portion of software maintenance efforts~\cite{8802807,922739}.
Traditionally, feature addition is manually performed by developers.
\emph{But what if new features could be added to a software project simply by updating its public documentation?}
In this setting, documentation updates act as executable, public-facing feature specifications that software engineering agents translate into code changes.
As software engineering agents are increasingly applied to repository-level development tasks, evaluating their ability to implement new features accurately becomes important.
We refer to the setting in which agents implement features from documentation updates as \textit{\paradigm}.
To support meaningful evaluation in this setting, high-quality benchmarks are essential, as they help assess agent capabilities, identify limitations, and guide future improvements.

Existing benchmarks for agentic software development primarily rely on issue reports as task inputs~\cite{jiang2025agentic}.
However, issue reports are typically maintainer-facing.
They are designed for coordination among developers~\cite{github_issues_feature}, are typically authored by individuals familiar with the codebase~\cite{got_issues}, and often include implementation-level technical details such as reproduction steps, stack traces, and test cases~\cite{chaparro2019assessing,soltani2020significance}.
In addition, issue-based benchmarks are focused primarily on bug fixing or issue resolution, with insufficient attention to feature addition.
Therefore, existing benchmarks do not fully capture the needs of evaluating \paradigm.

The key to evaluating \paradigm lies in identifying task inputs that reflect externally visible feature requirements.
Documentation provides a natural and effective source of such inputs.
It is written for external users and integrators of software projects~\cite{974401,alchimowicz2016coca} and serves as an authoritative description of software functionality and expected behavior.
Emerging development practices, such as README-driven development~\cite{prestonwerner_2010}, OpenAPI-based workflows~\cite{openapi_generator_2025}, and spec-driven development~\cite{github_spec_kit}, have followed a documentation-first manner.
As software evolves, documentation is often updated to reflect newly introduced or modified features~\cite{bennett2000software}.
These documentation changes specify intended behavioral changes in a public-facing form, making them a valuable source for constructing feature addition tasks.

Based on this observation, we introduce \nc, a benchmark for \paradigm.
\nc identifies feature addition tasks through release notes and constructs task inputs from \textit{changes in public-facing software documentation}, which serve as authoritative specifications of software functionality~\cite{aghajani2019software}.
It consists of 634 carefully collected tasks derived from 10 mature open-source software projects hosted on GitHub, spanning libraries, developer tools, and frameworks.
As illustrated in Figure~\ref{fig:task}, each instance takes documentation changes as input and expects a software engineering agent to generate corresponding code changes. The implementation is validated using developer-written test cases.
In addition, to provide a lightweight and reliable evaluation option, we manually validate a subset called \nc Verified. 
This subset consists of 114 instances with their task clarity and evaluation accuracy manually verified, enabling reliable evaluation under limited resources.

We leverage both \nc Verified and \nc Full to systematically evaluate a range of state-of-the-art software engineering agents with the Agentless~\cite{agentless} and OpenHands~\cite{wang2025openhands} scaffolds.
Evaluation results show that, despite incurring high token consumption, the best success rate is only 37.72\%, achieved when using Qwen3-Coder-480B with the OpenHands scaffold.
In comparison, when evaluated on SWE-bench, a benchmark collected from similar projects, the same model and scaffold achieve a success rate of 69.60\%. This stark performance gap highlights documentation-driven feature addition as a substantially more challenging and complementary benchmark setting.
Our analysis further reveals that software engineering agents face key challenges in performing cross-file edits, understanding existing code modules, and correctly interpreting documentation semantics.

In summary, our main contributions are:
\begin{itemize}[left=0em, topsep=0pt]
    \item We introduce \nc, a documentation-driven benchmark for feature addition in mature open-source software projects, using documentation diffs as public-facing feature specifications.
    \item We present a systematic five-phase construction pipeline to ensure the quality and fidelity of the benchmark, and further provide a manually validated subset for lightweight and reliable evaluation under limited resources.
    \item We comprehensively evaluate multiple state-of-the-art software engineering agents on \nc, analyze their performance from both quantitative and qualitative perspectives, and identify key factors that affect agent performance.
\end{itemize}

%% file: Figures/fig_task.tex
\begin{figure}[t]
    \centering
    \includegraphics[width=\columnwidth]{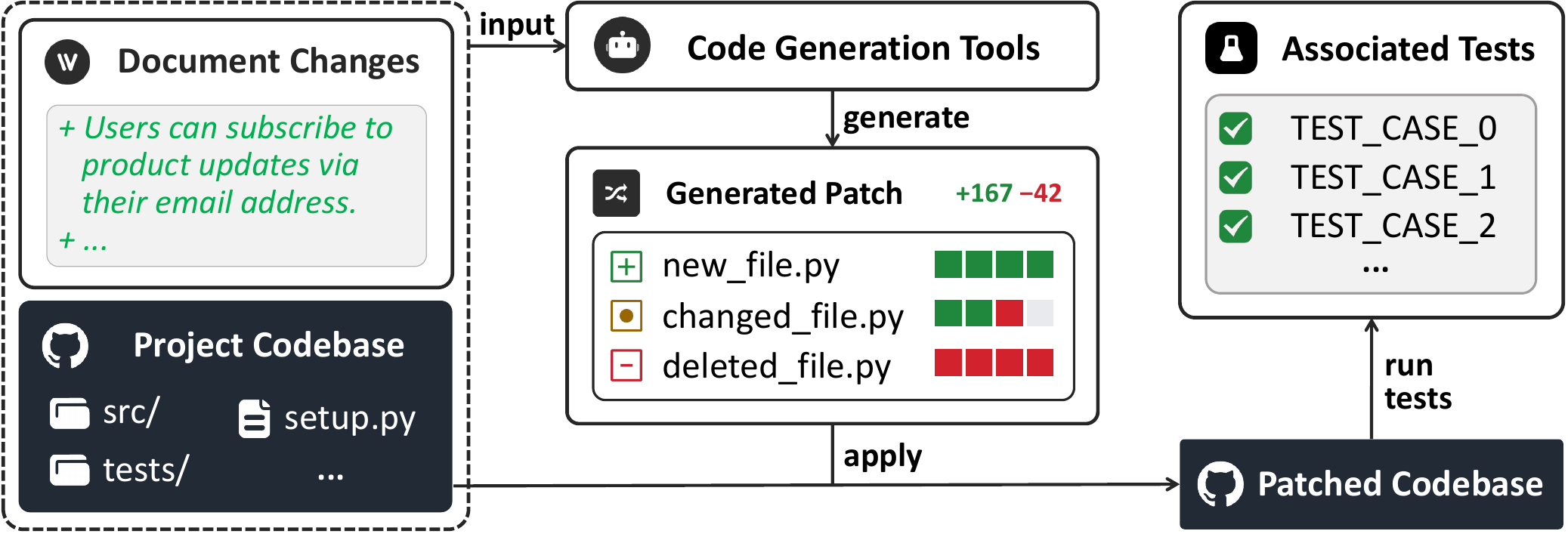}
    \caption{Overview of documentation-driven feature addition tasks.}
    \label{fig:task}
\end{figure}

%% file: Sections/02_featurebench.tex
\input{Figures/fig_framework}
\section{\nc}

\nc focuses on evaluating software engineering agents in documentation-driven feature-addition scenarios.
It comprises 634 tasks collected from 10 popular open-source projects. 
Each task corresponds to a real-world feature addition explicitly documented in the project's official release notes.
Each feature addition includes a coordinated change to the documentation, the code change, and the test suite.
The goal of each task is to produce a patch that implements the feature specified by the documentation change and passes the relevant tests.
The input includes the documentation changes and the complete codebase, as illustrated in Figure~\ref{fig:task}. 
This section outlines the construction process of \nc and its task formulation. In addition, we describe how we constructed a reliable subset, \nc Verified, through a combination of manual validation and refinement.

\subsection{Benchmark Construction}
We focus on high-quality, real-world software development data to construct \nc through the five phases shown in Figure~\ref{fig:framework}.

\subsubsection{Phase 1: Project Selection}
The project selection begins with the 12 projects used in SWE-bench~\cite{jimenez2024swebench}, which are actively maintained and well-documented open-source projects.
Unlike SWE-bench, which constructs instances by linking GitHub issues to corresponding pull requests (PRs), we identify real-world feature addition tasks from release notes, which are written or confirmed by project developers and typically include clearly labeled feature additions, making them a reliable source of accurate task annotations.
Well-maintained release notes often annotate each change with its type (e.g., ``feature'', ``enhancement'', ``bug fix'', etc.), making them a reliable source for identifying feature additions. 
In this context, we refer to each change item in a release note as an \emph{update entry}.
To ensure that the constructed benchmark is reliable, representative, and suitable for evaluating documentation-driven feature addition, we apply additional criteria when selecting projects. Specifically, we require that projects satisfy the following three conditions:
\begin{itemize}[left=0pt, topsep=0em]
    \item \textbf{Availability of release notes:} 
    Prior work~\cite{bi2020empirical,yu2009mining} shows that a project with release notes is more likely to be well-maintained.
    Moreover, release notes summarize code changes introduced in different versions, and thus are useful for identifying and collecting feature-adding instances.
    \item \textbf{Use of feature-labeled update entries:} 
    We define a \emph{feature addition} as a change that introduces new externally visible capabilities or extends existing functionality.
    Examples include adding a new button to trigger an action or allowing export into a new file format.
    In contrast, changes such as bug fixes, code refactoring, or performance tuning are not considered.
    We require that release notes explicitly label each entry with its type (e.g., ``feature'', ``enhancement'') to support reliable instance identification. 
    Without such labels, it is often unclear whether a code change corresponds to a feature addition or a bug fix.
    \item \textbf{Update entries linked to GitHub PRs:} 
    These links allow us to trace the documented features back to their corresponding code changes.
\end{itemize}
As illustrated in Phase 1 of Figure~\ref{fig:framework}, we can locate the release notes on the project's homepage.
The changelog for each version contains clearly labeled feature entries, each accompanied by a link to the corresponding PR (e.g. \colorbox[rgb]{0.51,0.74,0.57}{\textcolor{white}{Feature}} Added the objects. Text stat (\textcolor{blue}{\#3051})).
Following this approach, we evaluate all 12 projects from SWE-bench and find that 10 of them meet all the above conditions. The \texttt{flask}~\cite{flask} and \texttt{sympy}~\cite{sympy} projects lack feature-labeled annotations and are thus excluded.

\subsubsection{Phase 2: Instance Collection}
For each project selected in Phase 1, we retrieve all PRs corresponding to the feature additions identified from release notes, resulting in a total of 3,263 PRs. 
Then, we apply the following three filtering criteria:
\begin{itemize}[left=0pt, topsep=0em]
    \item \textbf{Modification of source code:} The PR must modify the source code of the project to ensure it implements the feature mentioned in the release note.
    \item \textbf{Modification of test files:} The PR must include changes to test files, ensuring that there are appropriate tests to verify the correctness of the feature implementation.
    \item \textbf{Modification of documentation:} The PR must update the project documentation, which can provide a clear description of the added feature.
\end{itemize}
Unlike SWE-bench, which considers only source code and test modifications, we additionally require documentation changes, which serve as the task specification provided as input. Compared to issue reports, documentation changes more directly describe intended feature behavior without exposing implementation details.
Additionally, we set a cutoff date of August 15, 2024, and retain only PRs merged before this date, aligning with SWE-bench for fair comparisons.
After filtering, we collect detailed information for each PR, including the \textit{base commit}, the \textit{documentation change}, \textit{feature\_patch}, and \textit{test\_patch}. 
The base commit refers to the commit on which the feature was originally developed.
A documentation change denotes the diff containing all changes made to the project documentation, which serves as the task input. 
\textit{feature\_patch} refers to the code change in the PR, which implements the new functionality;
\textit{test\_patch} is the test change in the PR, which is used to verify the correctness of the feature implementation.
Each PR that passes the filtering process is treated as an instance in our benchmark. 
In total, this phase yields 1,571 feature-adding instances that serve as candidates for subsequent processing.

\subsubsection{Phase 3: Environment Construction}
Unlike SWE-bench, which constructs a separate Docker image for each instance, we adopt a more resource-efficient strategy. 
SWE-bench's per-instance Docker approach incurs significant storage overhead and is less scalable when instances share highly similar environments~\cite{yang2025swesmith}. 
In our benchmark, many instances come from the same project and only differ slightly across minor versions, with occasional instance-specific requirements. 
Thus, we construct a single base Docker image per project and manage version isolation using Anaconda~\cite{anaconda2025}. 

To ensure the correct execution and evaluation of feature addition tasks, each instance must be built and executed in an isolated environment to prevent interference. 
In this phase, we rely on both automated identification and manual inspection to construct the runtime environment for each instance.
Specifically, for a given project, we first group instances by minor versions using automated scripts, which parse version-related files such as \textit{setup.py}, \textit{\_\_init\_\_.py}, and release notes. We then manually inspect the format and location of environment configuration files, such as \textit{README.md}, \textit{requirements.txt}, \textit{pyproject.toml}, or \textit{pom.xml} to identify how dependencies are declared. Automated scripts are developed to extract and install these dependencies accordingly.

Despite our careful exploratory setup, some instances cannot be fully built or executed due to issues such as loosely defined dependency versions or breaking updates in libraries. 
For example, if a project does not strictly pin the version of a dependency, an API used in the original code may be deprecated or changed in a newer library version, causing the instance to fail. 
To address this, we log the build and runtime outputs of each instance, and for the failed instances, we manually fix incorrect or missing dependency specifications and record the repair process as a reproducible \textit{shell} instruction. 
In practice, such manual refinement is required only occasionally.
Since instances within the same project often share similar environment problems, a single fix (e.g., adding or pinning the version of a missing dependency) typically applies to many instances, resulting in minimal additional human effort.
This phase ensures that each instance is equipped with a clean and functional containerized environment, establishing a reliable foundation for subsequent testing and analysis.

\subsubsection{Phase 4: Instance Filtering}
After collecting instances and constructing runnable environments, we proceed to verify each instance as a valid feature addition by analyzing the behavioral differences in the test execution before and after applying the feature patch. 
Specifically, we execute all test files that are modified or added in the \textit{test\_patch}, as they are expected to reflect the intended behavior of the new feature.
For each instance, we execute the relevant test cases and collect two sets of logs, i.e., \textit{before.log} and \textit{after.log}. \textit{before.log} is obtained by applying the \textit{test\_patch} to the base commit and then executing the corresponding test suite, while \textit{after.log} is obtained by applying both the \textit{test\_patch} and the \textit{feature\_patch} to the base commit, followed by test execution.

From these logs, we extract the execution status of each test case. Following the method used by SWE-bench, we focus primarily on the transition of test case status between before.log and after.log. Specifically:
\begin{itemize}[left=0pt, topsep=0em]
    \item We collect all test cases with a PASSED → PASSED transition as regression tests, and refer to them as \textit{P2P tests}. These test cases ensure that the addition of the new feature does not break existing features.
    \item To confirm that the feature is correctly implemented, we identify test cases with a FAILED → PASSED transition as validation tests, and refer to them as \textit{F2P tests}. These test cases validate that the added feature is now operational.
\end{itemize}
Together, these two types of test cases constitute the complete set of relevant tests for evaluating a given instance. 
If an instance does not have any F2P test, it is excluded from the benchmark, as there is no clear evidence that a feature has been added.
After applying this filtering, we retain a total of 634 instances.

It is important to note that, unlike SWE-bench, we do not discard instances in which before.log contains errors such as \textit{ImportError} or \textit{AttributeError}. These errors are often expected in feature addition scenarios because the new feature may introduce previously undefined modules, classes, or functions. As a result, test cases referring to these components will naturally fail before the feature patch is applied, and the corresponding instances should not be filtered when collecting feature addition tasks.
Among the 634 instances in \nc, 276 instances show such errors in before.log.

\input{Tables/table_verification}

\subsubsection{Phase 5: Input Refinement}
During benchmark construction, we observed that some feature additions introduce new program entities (e.g., files, classes, or methods) that are referenced in tests and implementation patches but not mentioned in the documentation change.
As the evaluation is test-driven, mismatched or missing entity names can cause models to fail tests even when the generated feature is correct. Simply discarding such cases, as done in prior work~\cite{jimenez2024swebench}, would remove legitimate feature additions and thus reduce benchmark coverage.

To preserve these instances while avoiding superficial evaluation failures,
we perform a structured static analysis to identify critical missing names.
Specifically, given an instance, we first switch the codebase to its base commit. 
We then identify the list of files modified by the \textit{feature\_patch} and the \textit{test\_patch}, denoted as \textit{feature files} and \textit{test files}, respectively. 
For each file list, we extract all file names and the identifiers (e.g., classes, functions, attributes) in these files through static analysis, producing a set of entities.
We refer to the entity sets extracted from feature files and test files as $F$ and $T$, respectively.
Next, we apply both \textit{feature\_patch} and \textit{test\_patch} to the codebase and repeat the same process to obtain the post-patch entity sets, denoted $F'$ and $T'$.
We compute the differences between the pre- and post-patch sets to obtain the sets of newly introduced entities in feature files and test files, respectively, i.e., $\Delta_F = F'-F$ and $\Delta_T = T'-T$.
We then compute the intersection of $\Delta_F$ and $\Delta_T$ to isolate entities that are newly introduced and referenced in both test and implementation code. 
This step ensures we only consider entities relevant to the validation process. 
Finally, we remove all entities that are already mentioned in the documentation changes, yielding the final set of undocumented but test-referenced entity names.
We refer to such entity names as \textit{identifier hints}, which 
are stored as auxiliary metadata and can optionally be appended to the model input during evaluation.
Accordingly, we provide two benchmark variants: one with \textit{identifier hints} and one without. This design allows users to choose whether to include such information, depending on their tolerance for potential data leakage.

In addition, we observed that some documentation changes include PR numbers, URLs, or references to issue trackers. 
These artifacts may guide models to recall memorized content or inspire models to access online resources, thus posing the risk of data leakage.
To mitigate this risk, we implement automated scripts to detect and mask such information in the documentation changes.

Together with previous phases, this yields the final set of 634 documentation-driven feature addition tasks used in \nc. 
A final breakdown of these task instances across projects is presented in Table~8 (see \appendixurl~A).

\subsection{\nc Verified}\label{sec:verified}
\nc collects a large number (634) of feature addition PRs from the wild, offering broad coverage and simulating real-world development scenarios.
However, evaluating software engineering agents on 634 tasks is expensive and time-consuming.
Furthermore, the specifications and the oracle (i.e., the test cases for our benchmark) of some real-world feature addition tasks are not well-specified, making them very challenging to tackle.
To provide a lightweight and dependable alternative for debugging, fine-grained analysis, and evaluation under limited resources, we construct a manually validated subset, namely \nc Verified.
We refer to the full set as \nc Full.
Together, the two sets offer complementary strengths, balancing broad coverage with lightweight evaluation.

\subsubsection{Construction Process}

\rev{To ensure that the verified subset is representative of the full set, it should include at least 84 instances, which is the minimum sample size required for 634 \nc Full instances at a 95\% confidence level with a 10\% error margin, following common practice in empirical software engineering~\cite{bhatia2020study, ahmad2017determining}.}
We perform random sampling, followed by manual verification to assess task quality.
To ensure project diversity, we \rev{draw a stratified random sample by project from \nc Full}.
Considering that low-quality samples may be filtered out during manual verification, we take up to 35 task instances per project, yielding a total of 238 candidate tasks.

Specifically, to carry out the annotations, we recruited five annotators, all of whom were screened for qualifications, including at least three years of experience with Python and a bachelor's degree or higher in a relevant field.
Following the annotation guidelines of the recently released SWE-bench Verified~\cite{openai2025swebenchverified}, each annotator is asked to label samples based on \textbf{the Clarity of the Task} and \textbf{the Accuracy of the Evaluation}.
Considering the subjectivity of human judgment, each sample is annotated independently by two different annotators.
For quality assurance, we conservatively select the more stringent label when disagreements arise, following SWE-bench Verified.
As a result, each annotator evaluated approximately 95 task instances.

\subsubsection{Annotation Criteria}
Our criteria used for manual annotation follow those of SWE-bench Verified, as follows:

\textbf{Clarity of the Task.}
Following SWE-bench Verified, we assess whether the task input (i.e., documentation changes in our work) clearly specifies the intended functionality. Annotators act as senior software engineers with full access to the codebase and related documentation. They judge whether the task is sufficiently well-defined to enable a meaningful implementation attempt and assign a clarity score accordingly. The scoring options range from 0 to 3, with lower scores indicating clearer and more precise documentation changes.

\textbf{Accuracy of the Evaluation.}
Unlike SWE-bench Verified, which focuses on whether tests can reliably evaluate diverse correct implementations (e.g., avoiding false negatives due to mismatches in naming or structure), we mitigate such concerns during benchmark construction in Phase 5. As a result, our evaluation emphasizes whether the provided test cases faithfully and comprehensively cover the functionality described in the documentation changes.
Annotators apply the test patch to the codebase and assess whether the added tests are appropriately scoped. Scoring is based on the criteria in Table~\ref{tab:verification}, with lower scores indicating more accurate and targeted coverage.

Additionally, if the annotator believes that there are other major problems with the sample that have not been considered, i.e., any other reason why this sample should not be used in our setup for evaluating coding ability, the annotator needs to flag the sample and provide a basic explanation.

\input{Tables/table_scores}
\subsubsection{Annotation Results}\label{sec:annotation-results}

After all the samples are annotated, we filter out any sample from the original test set where the severity of the clarity of the task or the accuracy of the evaluation is labeled as 2 or above. We also filter out all samples that have a major problem flagged by annotators. For example, in \textit{pytest-3576}, the associated PR includes a large number of unrelated documentation changes, which are not indicative of genuine feature additions and thus do not meet the criteria of our benchmark.
As summarized in Table~\ref{tab:scores}, a total of 106 instances were removed due to unclear task descriptions or inaccurate evaluations, and 18 instances were excluded due to other major problems.
\rev{The removed 106 instances consist of 77 instances flagged on task clarity and 73 on test accuracy, which overlap in 44 cases. Among the 18 excluded for other major problems, 3 were flagged unanimously by both annotators and thus left unscored along the two dimensions, so the Task Clarity and Test Accuracy columns in Table~\ref{tab:scores} total 235 rather than 238.}
Finally, we obtain 114 samples, which constitute \nc Verified.
During verification, Cohen's kappa is 0.4193, indicating moderate agreement and closely matching prior work (0.3906 on SWE-bench Verified~\cite{openai2025swebenchverified}), which supports the reliability of our process.
\rev{To further validate the reliability of the evaluation, we measured the line-level coverage of each golden feature patch achieved by its F2P tests. On \nc Verified, the median and mean coverage reach 94.60\% and 84.90\%, respectively, comparable to prior work (96.94\% and 72.95\% on SWE-bench Verified). This indicates that the retained F2P tests provide reliable evaluation signals.}

\subsubsection{\rev{Analysis of Filtered Instances}}\label{sec:annotation-insights}
\rev{To better understand why instances were filtered out during the construction of \nc Verified, two authors manually analyzed the prior annotation results, including annotator scores and explanations, together with each instance's documentation change, golden patch and test patch. This analysis covered the 77 instances filtered out under the task clarity dimension and the 73 filtered out under the evaluation accuracy dimension.
For each of the two dimensions, the two authors first jointly inspected 30 pilot instances to derive an initial coding scheme.
As the pilot annotation reveals a long-tail distribution, where the top-3 issue types cover 77\% and 71\% of the pilot instances along the two dimensions while each remaining type covers only 1.4 and 1.8 instances on average, we focus the analysis on the top-3 types in each dimension (shown in Table~\ref{tab:quality}).
Specifically, using this scheme, both authors independently categorized all filtered instances along the respective dimensions, reaching high agreement, with Cohen's kappa of 0.932 for task clarity and 0.913 for evaluation accuracy.
All disagreements were resolved through discussion.
}

\input{Tables/table_quality_issues}
\input{Figures/fig_annotation_example}

\rev{The upper part of Figure~\ref{fig:annotation-example} illustrates how task clarity issues arise from the gap between the documentation snippet and the implementation requirement.
In the \emph{changelog-only docs} example, the documentation only states that \texttt{Birch} supports \texttt{n\_clusters=None}, but the implementation requires knowing that this setting should bypass global clustering and directly form flat clusters from \texttt{\_CFSubcluster} objects.
In the \emph{non-feature changes} example, the documentation says that \texttt{LogisticRegression} is faster due to a new private loss-function module. The actual change is an internal optimization with no new user-facing API or behavior to implement.
In the \emph{scope mismatch} example, the documentation presents the change as adding a \texttt{while\_used} checker, while the PR also rewrites broader \texttt{while}-loop patterns across the codebase, so the documentation substantially understates the implementation scope.}

\rev{The lower part of Figure~\ref{fig:annotation-example} shows how evaluation accuracy issues arise when tests fail to exercise the intended behavior.
In the \emph{limited tests} example, the added test checks only one \texttt{FacetGrid} legend-order case for \texttt{pointplot}, leaving other label settings, hue interactions, and error cases uncovered.
In the \emph{adaptation-only tests} example, the patch repeatedly replaces local dependency checks such as \texttt{HAS\_PANDAS} with imports from \texttt{optional\_deps}. These changes adapt existing tests to a new import path but never validate the lazy access-time dependency checks introduced by the code patch.
In the \emph{missing integration tests} example, the tests use toy functions to exercise nested parameter validation in isolation, while the real integration points, such as estimator \texttt{fit()} methods and \texttt{non\_negative\_factorization}, remain untested.}

%% file: Figures/fig_framework.tex
\begin{figure*}[t]
    \centering
    \includegraphics[width=0.86\textwidth]{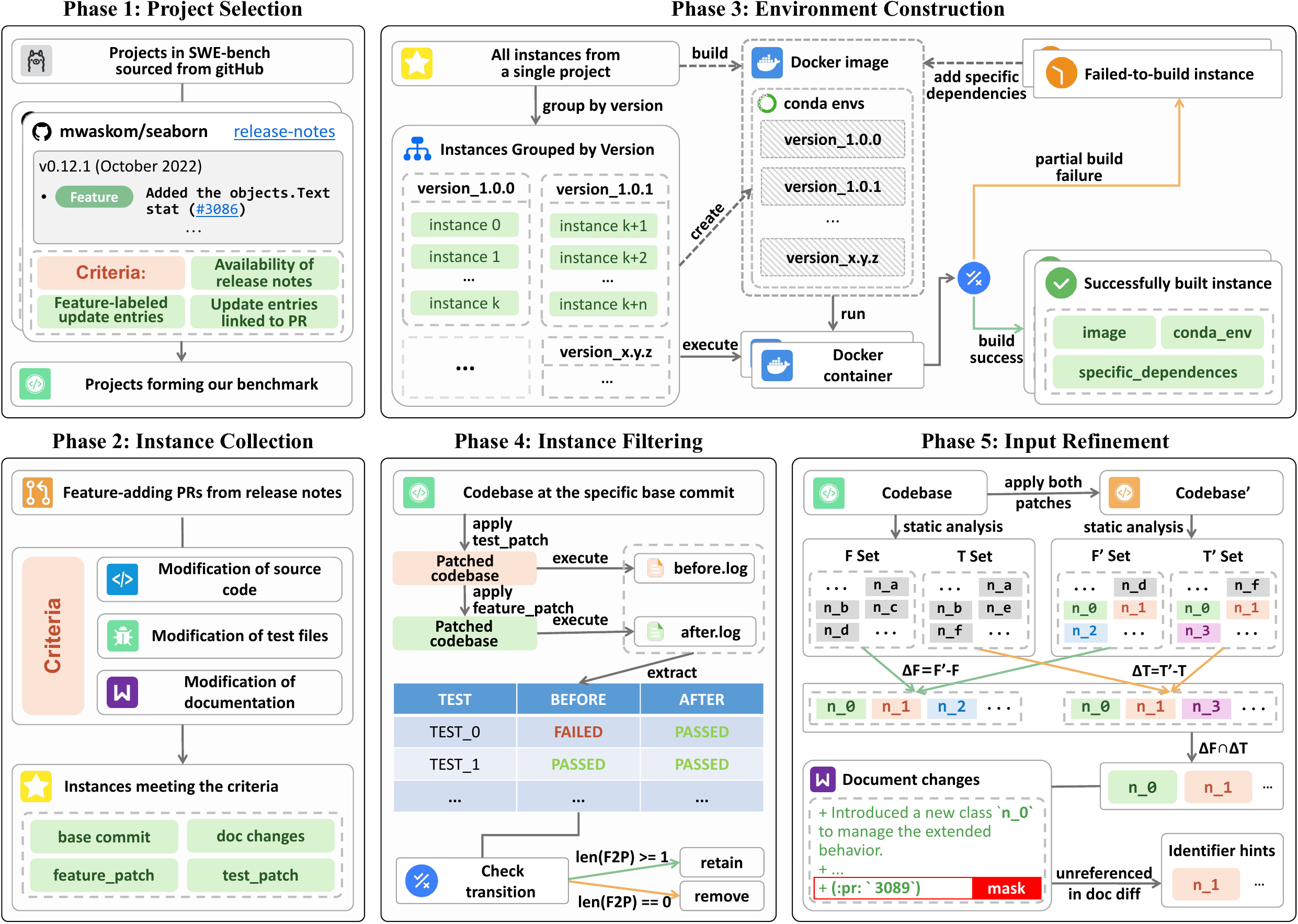}
    \caption{The workflow of building our benchmark.}
    \label{fig:framework}
\end{figure*}

%% file: Tables/table_verification.tex
\begin{table}
    \caption{Guidelines for the accuracy of the evaluation.}
    \begin{adjustbox}{width=\columnwidth}
    \begin{tabular}{cp{11cm}}
        \toprule
        \textbf{Score} & \textbf{Accuracy of the Evaluation} \\
        \midrule
        0 & The tests perfectly cover all aspects of the new feature mentioned in the doc changes. \\
        \midrule
        1 & The tests cover the majority of aspects of the feature involved in doc changes, but may overlook some special cases. \\
        \midrule
        2 & The tests are too narrow/broad, only test a few special cases, or test content far beyond the description in the document. \\
        \midrule
        3 & The tests look for something different from what the doc changes are about. \\
        \bottomrule
    \end{tabular}
    \end{adjustbox}
    \label{tab:verification}
\end{table}

%% file: Tables/table_scores.tex
\begin{table}
    \caption{Human verification results for task clarity and test accuracy. For the \textit{Other Problem} column, a score of 0 indicates no issues were identified, while a score of 3 indicates the presence of significant issues.}
    \centering
    \begin{adjustbox}{width=0.8\columnwidth}
    \begin{tabular}{lccc}
        \toprule
        Score & Task Clarity  & Test Accuracy & Other Problem \\
        \midrule
        0 (best) & 59 & 68 & 220 \\
        1 & 99 & 94 & -- \\
        2 & 59 & 64 & -- \\
        3 (worst) & 18 & 9 & 18 \\
        \bottomrule
    \end{tabular}
    \end{adjustbox}
    \label{tab:scores}
\end{table}

%% file: Tables/table_quality_issues.tex
\begin{table}[t]
    \centering
    \small
    \caption{Common quality issues in filtered instances.}
    \begin{adjustbox}{width=\columnwidth}
    \begin{tabular}{llp{0.65\columnwidth}}
        \toprule
        Type & Share & Description \\
        \midrule
        \multicolumn{3}{c}{\cellcolor{lb}\textbf{Task Clarity}} \\
        Changelog-only docs & 42.9\% & Brief changelog-only documentation without usage guidance, parameter semantics, or examples. \\
        Non-feature changes & 19.5\% & Internal refactoring, performance optimization, or deprecation changes without user-visible functionality. \\
        Scope mismatch & 15.6\% & Implementation scope substantially exceeds the behavior specified by the documentation diff. \\
        \midrule
        \multicolumn{3}{c}{\cellcolor{lb}\textbf{Evaluation Accuracy}} \\
        Limited tests & 30.1\% & Minimal tests covering only basic functionality, without boundary conditions (e.g., empty or extreme inputs, alternative parameter settings) or error cases. \\
        Adaptation-only tests & 21.9\% & Test changes that mainly adapt identifiers or imports without validating new behavior. \\
        Missing integration tests & 16.4\% & Unit tests for a new module without integration coverage for modified existing files. \\
        \bottomrule
    \end{tabular}
    \end{adjustbox}
    \label{tab:quality}
\end{table}

%% file: Figures/fig_annotation_example.tex
\begin{figure}[t]
    \centering
    \includegraphics[width=\columnwidth]{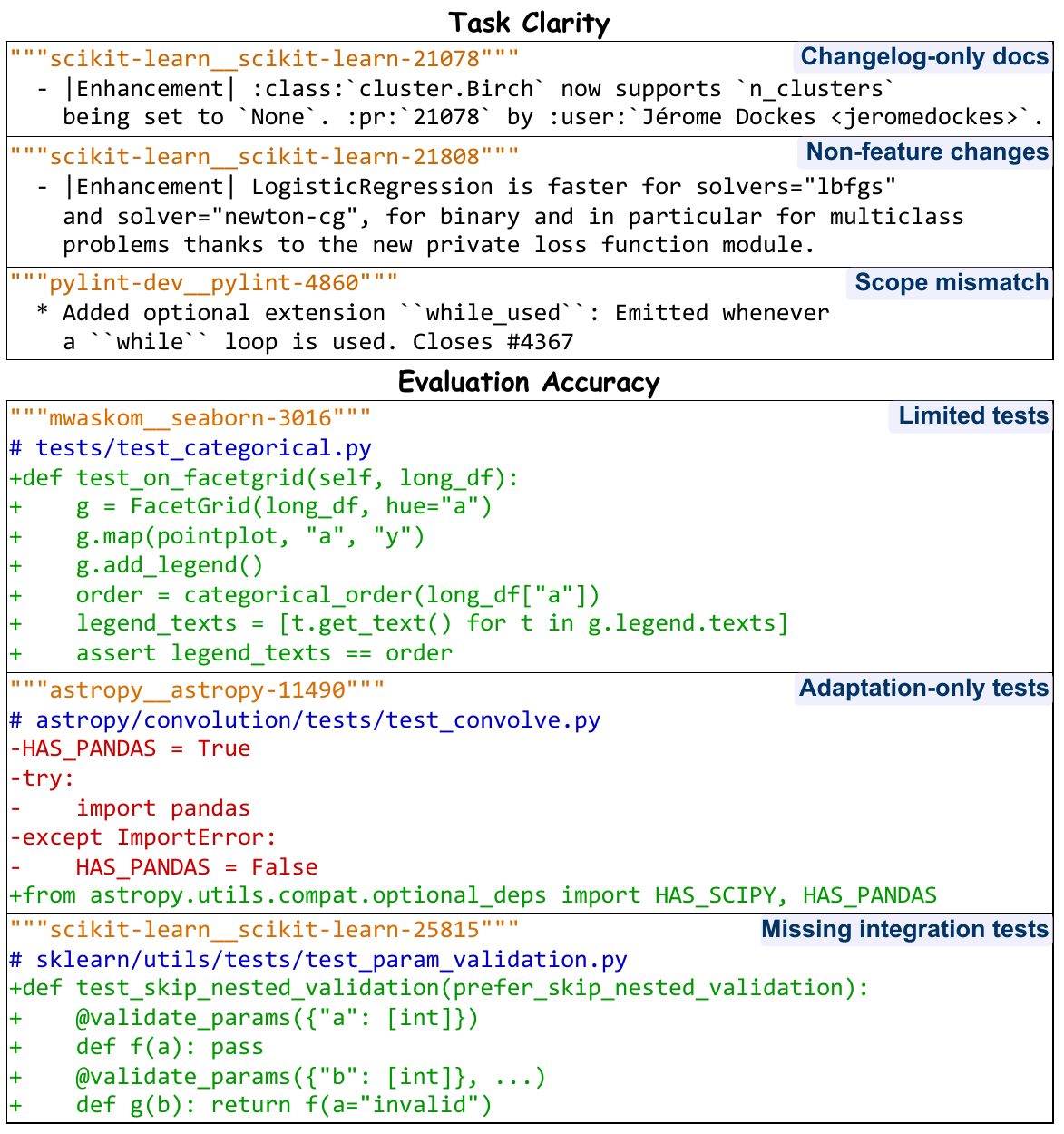}
    \caption{Representative examples of task-clarity and evaluation-accuracy issues in filtered instances.}
    \label{fig:annotation-example}
\end{figure}

%% file: Sections/03_benchmark_stat.tex
\section{Benchmark Characteristics}\label{Benchmark_Characteristics}

\input{Figures/fig_challenge}

To better understand unique characteristics and challenges posed by \nc, we analyze the distribution of instances in \nc and \swe across dimensions related to input complexity, localization difficulty, and editing difficulty, as shown in Figure~\ref{fig:patch-distributions}. 
For visualization purposes only, we omit extreme values (the top 1\% of values in each dimension) from Figure~\ref{fig:patch-distributions} to prevent distortion; these values are included in all statistical analyses.

\textbf{Input complexity.}
Unlike \swe, which uses issue descriptions as input, \nc takes documentation changes as input and requires software engineering agents to understand the semantic differences before and after the change.
Long documentation changes pose extra challenges for agents to comprehend the task and generate appropriate outputs.
As shown in Figure~\ref{fig:patch-distributions}(a), the average length of issue descriptions in \swe Full and Verified is 480.37 and 447.72, respectively, while the average length of documentation changes in \nc Full and Verified is 739.06 and 820.65, which is nearly twice as long as that in \swe.
This indicates that tasks in \nc require agents to capture information from longer and more complex task inputs, posing greater challenges to long-text comprehension, identifying key information, and attending to fine-grained details.

\textbf{Localization Difficulty.}
Accurate localization in the codebase is a necessary condition for generating the correct patches. 
Localization is particularly challenging when multiple files are involved or when the changes are scattered across many code regions.
As shown in Figure~\ref{fig:patch-distributions}(b,c), on average, golden patches in \nc Full involve edits to 2.65 files and 10.32 hunks, whereas in \swe Full, they involve only 1.66 files and 4.12 hunks.
In the Verified subsets, the number of files involved is 2.39 for \nc and 1.24 for \swe, with corresponding hunk counts of 8.96 and 2.43.
Moreover, Figure~\ref{fig:patch-distributions}(d) reveals that \nc includes a much higher proportion of patches that involve adding or deleting files. 
13.56\% of instances in \nc Full require new files, 
while the percentage is only 1.70\% in \swe Full.
These results indicate that our benchmark requires agents to locate multiple code blocks spread across different files within large codebases, which goes far beyond the demands of issue-solving tasks in \swe, leading to localization complexity.

\textbf{Editing Difficulty.}
Once the correct edit location is identified, the correct code segments must be generated to realize the intended functionality. 
Larger edits typically increase the chance of introducing syntactic or semantic errors.
We measure the total number of lines changed in golden patches, i.e., the ground truth edits excluding changes to tests and documentation files, as shown in Figure~\ref{fig:patch-distributions}(e). 
In \nc Full, golden patches involve an average of 179.12 lines of edits, and 179.22 lines in \nc Verified.
In contrast, golden patches in \swe Full average 37.71 lines of edits, and only 14.32 lines in \swe Verified.
Furthermore, we find that nearly all \swe patches are under 200 lines, while almost 20\% of \nc patches exceed that length.
These larger modifications may require stronger code generation capabilities and greater coherence across multi-line edits.

%% file: Figures/fig_challenge.tex
\begin{figure*}
  \centering
  \includegraphics[width=\textwidth]{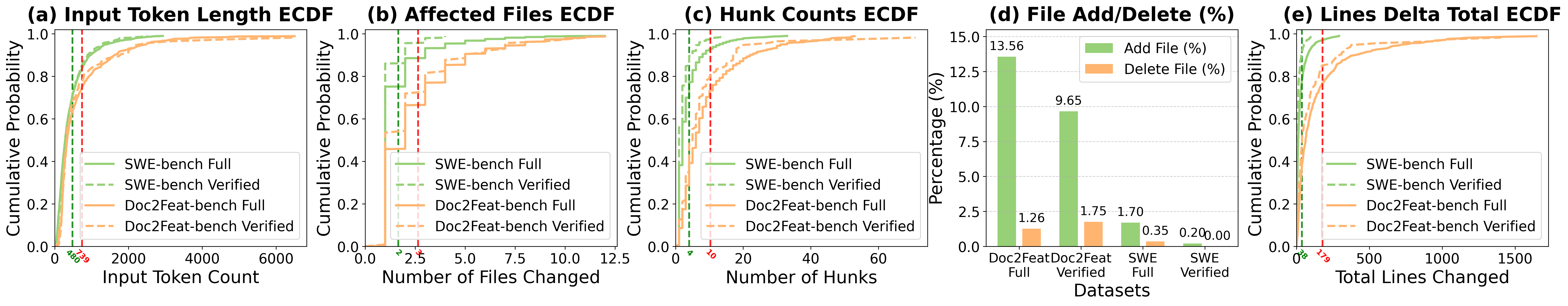}
  \caption{
    Comparison of golden patch distributions between \nc and \swe along dimensions related to localization and editing difficulty. ECDF stands for ``empirical cumulative distribution function''.}
  \label{fig:patch-distributions}
\end{figure*}

%% file: Sections/04_setting.tex
\section{Experimental Setup}

\subsection{Research Questions}

\begin{itemize}[left=0pt, topsep=0em]
    \item RQ1: How do state-of-the-art software engineering agents perform on \nc Verified?
    \item RQ2: How do state-of-the-art software engineering agents perform on \nc Full?
    \item RQ3: What are the reasons for agent failures on \nc?

\end{itemize}

\subsection{Model Selection}
Completing a task in \nc typically requires a deep understanding of the target project, demanding strong reasoning ability and the capacity to handle long contexts.
Considering these requirements, we selected 8 state-of-the-art models for evaluation on \nc Verified.
Four of them are open-source models, i.e., DeepSeek-v3-0324~\cite{deepseekv3}, DeepSeek-R1-0528~\cite{deepseek-r1}, Qwen3-235B-A22B~\cite{qwen3}, and Qwen3-Coder-480B-A35B, where DeepSeek-R1 is a reasoning model and Qwen3-235B-A22B is used in thinking mode.
The other four are closed-source models, i.e., Claude-4-Sonnet~\cite{claude4}, GPT-4o-2024-11-20~\cite{openai2025gpt4o}, Gemini-2.5-Pro~\cite{gemini2.5}, and GPT-5-mini~\cite{OpenAI2025GPT5}, where Gemini-2.5-Pro is a reasoning model and GPT-5-mini is used in medium thinking.
For brevity, version numbers will be omitted when referencing model names throughout the paper.
\nc Full is significantly larger than \nc Verified, with 5.6 times the number of instances.
Due to the high cost of Claude-4-Sonnet, Qwen3-Coder, GPT-5-mini, and Gemini-2.5-Pro (total estimated at USD\,5,124) running on \nc Full, we focus on \nc Verified for those models.

\input{Tables/result_verified}

\subsection{Scaffolds}
Due to the complexity of documentation-driven feature addition tasks, directly prompting models to generate the required changes is impractical.
Therefore, we adopt two state-of-the-art general scaffolds for solving software engineering tasks: a pipeline-based method, Agentless~\cite{agentless}, and an agent-based method, OpenHands~\cite{wang2025openhands}.
They respectively represent two distinct paradigms to automate SE tasks with software engineering agents, and both have attained top scores on \swe.

\begin{itemize}[left=0pt, topsep=0em]

\item\textbf{Agentless} adopts a hierarchical approach to sequentially identify the relevant files, classes or methods, and lines of code that require modification, and then generates patches based on the localized content with the selected model.
Following Guo et al.~\cite{guo2025omnigirl} and Zan et al.~\cite{zan2025multiswebench}, we use Agentless-1.0 for evaluation.

\item\textbf{OpenHands} is a widely adopted platform for building software development agents.
We adapt the documentation changes in \nc as inputs to the OpenHands CodeAct Agent and utilize the standard evaluation Docker images provided by \nc as a sandbox environment to ensure the correct execution of tasks.
\end{itemize}

\subsection{Evaluation Metrics} 
Following prior studies~\cite{jimenez2024swebench, zan2025multiswebench, guo2025omnigirl}, our end-to-end evaluation primarily considers the success rate (i.e., the proportion of instances for which the given patch passes the accompanying test suite) of feature addition tasks (\textbf{Success\%}), the success rate of patch application (\textbf{Applied\%}), and the average token cost (\textbf{\#Token}). 
Additionally, two other metrics are used: \textbf{File Matched Rate (File\%)} measures alignment between model-edited and reference files, and \textbf{Regression Tests Pass Rate (RT\%)} assesses the preservation of existing functionality. 

We introduce a new metric named \textbf{Feature Validation Rate (FV)} to evaluate the completeness of an agent's implementation of a specific feature, which we measure based on the pass rate of F→P tests.
It is reported at both micro and macro levels to reflect overall test-level performance and instance-level robustness, respectively.

\begin{itemize}[left=0pt, topsep=0em]
    \item \textbf{Micro-level FV (FV-Micro):}
    The overall ratio of passed F→P tests to the total number of F→P tests across all instances: \\
    $\text{FV-Micro} = \frac{\sum_{i=1}^{N} \text{\#Passed}_i}{\sum_{i=1}^{N} \text{\#Total}_i}$
    where $\text{\#Passed}_i$ and $\text{\#Total}_i$ denote the number of passed and total F→P tests for instance $i$, respectively. $N$ is the total number of instances.

    \item \textbf{Macro-level FV (FV-Macro):}
    The average F→P pass rate across instances:
    $\text{FV-Macro} = \frac{1}{N} \sum_{i=1}^{N} \frac{\text{\#Passed}_i}{\text{\#Total}_i}$.

\end{itemize}

%% file: Tables/result_verified.tex
\begin{table*}[t]
\centering
\caption{Performance comparison across different software engineering agents on \nc Verified.
}
\begin{adjustbox}{width=1\textwidth}
\setlength{\tabcolsep}{3pt}
\begin{tabular}{lrrrrrrr|rrrrrrr}
\toprule
\multirow{2}{*}{\textbf{Model}}& \multicolumn{7}{c}{\textbf{Agentless}}  & \multicolumn{7}{c}{\textbf{OpenHands}} \\
\cmidrule{2-15}
& \textbf{Success\%} & \textbf{Applied\%} & \textbf{RT\%} & \textbf{FV-Micro} & \textbf{FV-Macro} & \textbf{File\%} & \textbf{\#Token} & \textbf{Success\%} & \textbf{Applied\%} & \textbf{RT\%} & \textbf{FV-Micro} & \textbf{FV-Macro} & \textbf{File\%} & \textbf{\#Token} \\
\midrule
\multicolumn{15}{c}{\cellcolor{lb} \textbf{Open-Source Model}} \\
\midrule
\textbf{Qwen3-235B}& 13.16& \textbf{100.00} & 76.32& 8.75\%& 22.39\%& 42.98 & 0.12M& 7.89& 64.91& 47.37& 1.96\%& 14.03\%& 45.61 &0.24M\\
\textbf{DeepSeek-v3}& 21.05& 98.25 & 78.95& 7.96\%& 32.80\%& 57.14 & 0.30M& 11.40& 65.79& 49.12& 1.68\%& 18.29\%& 56.14 &0.59M\\
\textbf{DeepSeek-R1}& 25.44& 93.86 & 73.68& \textbf{10.87\%}& 35.52\%& 50.47 & 0.50M& 7.02& 47.37& 46.49& 0.47\%& 10.86\%& 38.60 &0.39M\\
\textbf{Qwen3-Coder-480B}& \textbf{28.07}& \textbf{100.00} & \textbf{82.46}& 9.03\%& \textbf{40.12\%}& 54.39 & 0.45M& \textbf{37.72}& \textbf{92.11}& \textbf{84.21}& 10.96\%& \textbf{50.45\%}& 63.16 &2.91M\\
\midrule
\multicolumn{15}{c}{\cellcolor{lb} \textbf{Closed-Source Model}} \\
\midrule
\textbf{GPT-4o}& 13.16& \textbf{100.00}& 67.54& 7.97\%& 22.32\%& 47.37 & 0.11M& 5.26& 63.16& 57.89& 0.45\%& 6.76\%& 34.21 &0.58M\\
\textbf{Gemini-2.5-Pro}& 12.28& \textbf{100.00}& 74.56& 6.22\%& 20.55\%& 48.25 & 0.29M& 0.00 & 54.39& 61.40& 0.01\%& 0.29\%& 0.00 &0.47M\\
\textbf{Claude-4-Sonnet}& \textbf{28.07}& \textbf{100.00}& 79.82& 8.47\%& 38.48\%& \textbf{57.89} & 0.33M& 25.44& 68.42& 69.30& 11.25\%& 36.48\%& \textbf{67.54} &2.32M \\
\textbf{GPT-5-mini}& 26.32& 98.25& \textbf{82.46}& 10.63\%& 37.82\%& 53.98 & 0.21M& 36.84& 87.72& 80.70& \textbf{12.61\%}& 49.64\%& 64.60 & 1.69M\\
\midrule\midrule
\textbf{Average} 
& 20.94 & 98.80 & 76.97 & 8.74\% & 31.25\% & 51.56 & 0.29M
& 16.45 & 67.98 & 62.06 & 4.92\% & 23.35\% & 46.23 & 1.15M \\
\bottomrule
\end{tabular}
\end{adjustbox}
\label{verified}
\end{table*}

%% file: Sections/05_results.tex
\section{Results}

\subsection{RQ1: Performance on \nc Verified}

The evaluation results are shown in Table~\ref{verified}.
We find that even the most advanced software engineering agents still exhibit limited performance in documentation-driven feature addition.
Specifically, among the evaluated models, Qwen3-Coder-480B achieves the highest success rate of 37.72\% on \nc Verified with the OpenHands scaffold. 
The closed-source models GPT-4o and Gemini-2.5-Pro exhibit lower scores, with success rates of 5.26\% and 0\%. 
These results highlight significant performance differences across models of varying sizes and architectures under the same scaffold. 
Notably, the success rate on \nc Verified is much lower than that on \swe Verified.
For example, Claude-4-Sonnet combined with OpenHands achieves 70.4\% on \swe Verified but only 25.44\% on \nc Verified, implying that documentation-driven feature addition tasks present a substantially greater challenge compared to issue resolution tasks.

The Agentless scaffold achieves an average patch application rate of 98.80\%, whereas OpenHands produces applicable patches for only 67.98\% of the tasks on average.
For regression tests, 76.97\% of the patches generated by Agentless passed all regression tests on average, compared to only 62.06\% for OpenHands.
In terms of full F2P test suite pass rate, Agentless achieves 8.74\% (FV-Micro), with an average of 31.25\% (FV-Macro) of F2P tests passed per instance, while OpenHands achieves 4.92\% and 23.35\%, respectively.
The results show that Agentless outperforms OpenHands on \nc Verified.
Through manual inspection, we find that the reason lies in how OpenHands maintains a dialogue history to record interactions between the agent and the repository. Since solving tasks in \nc Verified requires editing an average of 2.39 files, the observed context often exceeds the model's context window, resulting in a large number of incomplete or empty patches.
In contrast, Agentless uses hierarchical localization to limit the length of retrieved context and applies post-processing to verify patch completeness, leading to better overall performance.

Notably, Gemini-2.5-Pro with OpenHands solved no tasks on \nc, largely because it failed to produce valid tool invocation formats, preventing repository access and modification. 
We further discuss this issue in \appendixurl~B.

\answerRQ{
State-of-the-art software engineering agents perform poorly on documentation-driven feature addition tasks. 
Even the best-performing model, Qwen3-Coder-480B with OpenHands, achieves only 37.72\% success rate on \nc Verified, while GPT-4o and Gemini-2.5-Pro achieve 5.26\% and 0\%, respectively.
This demonstrates that \nc introduces a novel and challenging task that enables the evaluation of agentic SE capabilities from a unique perspective.
}

\subsection{RQ2: Performance on \nc Full}\label{sec:rq2}
\input{Tables/result_full}

Considering the better average performance of Agentless and the high cost of OpenHands, we use Agentless to evaluate the models on \nc Full, and report the results in Table~\ref{full}.
We find that all four models show decreased performance on \nc Full compared to \nc Verified.
Specifically, the average success rate of the four used models drops from 18.20\% to 10.65\%, and the file matched rate drops from 49.49\% to 43.85\%. 
This drop may be primarily due to the increased scale and complexity, as shown in Section~\ref{Benchmark_Characteristics}, as well as the presence of some noise (e.g., ambiguous documentation changes or overly specific tests) in \nc Full.
Since \nc Full is a large-scale dataset involving more extensive file modifications and more complex code changes, it presents a diverse range of challenges.
In addition, \nc Full may contain some noise, including unclear documentation changes or low-coverage regression tests, which make it more representative of real-world documentation-driven feature addition tasks, but harder for agents to solve.
These factors make it more challenging.

It is worth noting that the ranking and relative performance of the models on \nc Verified are preserved on \nc Full except for Qwen3-235B and GPT-4o. 
For DeepSeek-R1, DeepSeek-v3 and Qwen3-235B, success rates drop from 25.44\% to 14.83\%, 21.05\% to 11.83\%, and 13.16\% to 6.62\%, respectively, while GPT-4o only drops from 13.16\% to 9.31\%. 
Qwen3-235B and GPT-4o achieve similar performance on \nc Verified, while GPT-4o outperforms Qwen3-235B on \nc Full.
These results demonstrate that compared to DeepSeek-v3 and Qwen3-235B, GPT-4o performs more robustly in the scenarios that are closer to real-world conditions.
A larger decline of Qwen3-235B than GPT-4o suggests that model robustness can vary as the evaluation shifts from a curated, high-quality subset to a larger set in real-world settings.

\answerRQ{
On \nc Full, DeepSeek-R1 achieves the best success rate of 14.83\%. 
The average success rate drops from 18.20\% on \nc Verified to 10.65\% on \nc Full, showing that \nc Full is substantially more challenging. 
Most models preserve similar relative rankings across the two datasets, indicating broadly consistent capabilities across cleaner and more realistic settings. 
GPT-4o is a notable exception, showing stronger robustness by outperforming Qwen3-235B on \nc Full despite similar performance on \nc Verified.
}

\subsection{RQ3: Failure Analysis}\label{sec:failure_analysis}

To investigate why agents fail to solve tasks on \nc, we manually examine 160 failed instances (randomly sampling 10 instances per model per scaffold, i.e., 10 × 2 × 8 = 160) and identify the following primary reasons for failures on \nc through a qualitative analysis of execution traces.

\textbf{Lack of cross-file editing capability.}
In \nc Verified and \nc Full, only 53.5\% and 45.7\% of the instances can be solved by editing a single file, respectively, indicating that \nc requires strong cross-file editing capability.
To further investigate how cross-file editing capability affects performance on feature addition tasks, we compute the number of modified files in the golden patch for each instance and divide the dataset into two groups: \textit{single-file} (\#modified\_files = 1) and \textit{multi-file} (\#modified\_files > 1). 
We analyze the performance of software engineering agents on \textit{single-file} and \textit{multi-file} tasks, and present the results in Table~\ref{table:CrossFileEval}.
On \nc Verified and \nc Full, the best-performing model solves only 20.75\% and 6.40\% of the \textit{multi-file} instances, respectively.
These results indicate that the lack of cross-file editing capability is one of the main reasons for failures on \nc.

\input{Tables/cross_file}

\textbf{Lack of comprehensive understanding of existing code modules.} 
Through manual inspection, we find that agents often attempt to implement new features by directly modifying the code of existing features, and 39.38\% of the examined instances exhibit this error.
This can lead to the new feature overriding existing features, resulting in failure in regression tests.
A typical example is the patch generated by DeepSeek-v3+Agentless on the task ``\textit{pylint-7869}''.
Figure~\ref{fig:understand} highlights a critical difference between the patch generated by DeepSeek and the golden patch for the task ``pylint-7869''.
This task requires introducing a new ``no-header'' option to Pylint. 
When this option is enabled, Pylint's output will omit (i.e., not display) the header information that normally shows the name of the module being analyzed.
DeepSeek-v3 directly modifies the member functions of the \emph{ColorizedTextReporter} class to implement the ``no-header'' option, whereas the golden patch introduces a new \emph{NoHeaderReporter} class to achieve the same goal.
Compared to the golden patch, DeepSeek-v3's modification breaks the original functionality. 
Although it passes all F2P tests, it results in numerous regression test failures.
This example demonstrates how agents may lack understanding of existing code modules needed to extend functionality without breaking existing features.
Users may alleviate such phenomena by providing the agent with information about the differences between existing modules or by guiding the agent to analyze existing modules through a predefined plan.

\input{Figures/fig_understand}

\textbf{Documentation semantic misinterpretation.}
\input{Figures/fig_understand_doc}
To understand how models misinterpret documentation semantics, we identify three of the most frequent misinterpretation patterns during manual inspection, and illustrate each of them with a representative case in Figure~\ref{fig:semantic_misunderstanding}.
\ding{182} The model misunderstands when/how exceptions should be raised or suppressed (33.13\%). 
For example, in \textit{astropy-11691}, the documentation requires invalid attribute-style access to follow Python's module-level \texttt{\_\_getattr\_\_} behavior~\cite{python_model} and raise \texttt{AttributeError}, while the model only extends name-based lookup logic and fails to fully preserve the intended exception semantics.
\ding{183} The model misinterprets edge conditions, thresholds, or conditional clauses (20.00\%). 
For example, in \textit{astropy-11954}, the documentation specifies strict constraints on \texttt{keys\_left}/\texttt{keys\_right}, such as mutual exclusivity with \texttt{keys} and consistency requirements, but the model mainly treats the feature as a temporary-key transformation and does not fully implement the documented parameter contract.
\ding{184} The documentation does not specify how to handle invalid input, and the implementation generated by the model differs from that of the developers (7.88\%). 
For example, in \textit{matplotlib-18715}, although the documentation requires the new configuration option to preserve the original behavior when set to \texttt{auto}, the model changes the validation boundary and thus deviates from the developers' intended default behavior.

\answerRQ{Failures on \nc mainly stem from three issues. First, agents struggle with cross-file editing, leading to very low success rates on multi-file tasks. Second, they often lack a clear understanding of existing code modules and may break prior functionality when adding features. Third, they frequently misinterpret documentation semantics, such as exception handling and edge conditions.
}

%% file: Tables/result_full.tex
\begin{table}[t]
\centering
\caption{Performance comparison across different software engineering agents on \nc Full with Agentless.}
\begin{adjustbox}{width=\columnwidth}
\setlength{\tabcolsep}{3pt}
\begin{tabular}{lrrrrrrr}
\toprule
\textbf{Model} & \textbf{Success\%}    &\textbf{Applied\%}   
& \textbf{RT\%} & \textbf{FV-Micro}   & \textbf{FV-Macro}  & \textbf{File\%} & \textbf{\#Token}    \\
\midrule
\textbf{Qwen3-235B}& 6.62&99.37& 75.24& 10.78\%& 15.77\%& 39.46& 0.19M\\
\textbf{DeepSeek-v3}& 11.83&\textbf{99.84}&  77.44& \textbf{16.02\%}& 24.09\%& \textbf{47.48}& 0.31M\\
\textbf{DeepSeek-R1}& \textbf{14.83}&95.90&  \textbf{79.34}& 10.53\%& \textbf{25.78\%}& 44.89& 0.57M\\
\textbf{GPT-4o}& 9.31&97.00& 71.14& 8.68\%& 17.81\%& 43.58& 0.11M\\
\midrule\midrule
\textbf{Average}& 10.65& 98.03& 75.79& 11.50\% & 20.86\% & 43.85& 0.30M\\

\bottomrule
\end{tabular}
\end{adjustbox}
\label{full}
\end{table}

%% file: Tables/cross_file.tex
\begin{table}[t]
\centering
\caption{Performance across different software engineering agents on \nc for \textit{single-file} and \textit{multi-file} tasks.}
\label{table:CrossFileEval}
\begin{adjustbox}{width=\columnwidth}
\setlength\tabcolsep{4pt}
\begin{tabular}{llcccc}
\toprule
\multirow{2}{*}{\textbf{Model}} & \multirow{2}{*}{\textbf{Scaffold}} & \multicolumn{2}{c}{\textbf{Single-File Modification}} & \multicolumn{2}{c}{\textbf{Multi-File Modification}} \\
\cmidrule(r){3-4} \cmidrule(r){5-6}
& & \textbf{Success\%} & \textbf{Applied\%} & \textbf{Success\%} & \textbf{Applied\%} \\
\midrule

\multicolumn{6}{c}{\cellcolor{lb} \textbf{\nc Verified}} \\
\midrule
 \multicolumn{2}{c}{\textbf{Total Instances}}& \multicolumn{2}{c}{61}& \multicolumn{2}{c}{53}\\
\midrule

\multirow{2}{*}{\textbf{Qwen3-235B}} 
& Agentless & 21.31 (13) & \textbf{100.00 (61)} & 3.77 (2) & \textbf{100.00 (53)} \\
& OpenHands  & 13.11 (8) & 65.57 (40) & 1.89 (1) & 64.15 (34) \\

\midrule
\multirow{2}{*}{\textbf{DeepSeek-v3}} 
& Agentless & 31.15 (19) & \textbf{100.00 (61)} & 9.43 (5) & 96.23 (51) \\
& OpenHands  & 14.75 (9) & 67.21 (41) & 7.55 (4) & 64.15 (34) \\

\midrule
\multirow{2}{*}{\textbf{DeepSeek-R1}} 
& Agentless & 40.98 (25) & 98.36 (60) & 7.55 (4) & 88.68 (47) \\
& OpenHands  & 11.48 (7) & 47.54 (29) & 1.89 (1) & 47.17 (25) \\

\midrule
\multirow{2}{*}{\textbf{Qwen3-Coder-480B}} 
& Agentless 
& 40.98 (25) & \textbf{100.00 (61)} 
& 13.21 (7)  & \textbf{100.00 (53)} \\
& OpenHands  
& \textbf{52.46 (32)} & 93.44 (57) 
& \textbf{20.75 (11)} & 90.57 (48) \\

\midrule
\multirow{2}{*}{\textbf{GPT-4o}} 
& Agentless & 19.67 (12) & \textbf{100.00 (61)} & 5.66 (3) & \textbf{100.00 (53)} \\
& OpenHands  & 6.56 (4) & 68.85 (42) & 3.77 (2) & 56.60 (30) \\

\midrule
\multirow{2}{*}{\textbf{Gemini-2.5-Pro}} 
& Agentless & 19.67 (12) & \textbf{100.00 (61)} & 3.77 (2) & \textbf{100.00 (53)} \\
& OpenHands  & 0.00 (0) & 52.46 (32) & 0.00 (0) & 56.60 (30) \\

\midrule
\multirow{2}{*}{\textbf{Claude-4-Sonnet}} 
& Agentless & 44.26 (27) & \textbf{100.00 (61)} & 9.43 (5) & \textbf{100.00 (53)} \\
& OpenHands  & 39.34 (24) & 68.85 (42) & 9.43 (5) & 67.92 (36) \\

\midrule
\multirow{2}{*}{\textbf{GPT-5-mini}} 
& Agentless 
& 39.34 (24) & 96.72 (59) 
& 11.32 (6)  & \textbf{100.00 (53)} \\
& OpenHands  
& \textbf{52.46 (32)} & 91.80 (56) 
& 18.87 (10) & 83.02 (44) \\

\midrule
\multicolumn{6}{c}{\cellcolor{lb} \textbf{\nc Full}} \\
\midrule

 \multicolumn{2}{c}{\textbf{Total Instances}}& \multicolumn{2}{c}{290}& \multicolumn{2}{c}{344}\\
\midrule

\multirow{1}{*}{\textbf{Qwen3-235B}} 
& Agentless & 11.38 (33) & 99.66 (289) & 2.62 (9) & 99.13 (341) \\

\midrule
\multirow{1}{*}{\textbf{DeepSeek-v3}} 
& Agentless & 20.00 (58) & \textbf{100.00 (290)} & 4.94 (17) & \textbf{99.71 (343)} \\

\midrule
\multirow{1}{*}{\textbf{DeepSeek-R1}} 
& Agentless & \textbf{24.83 (72)} & 96.55 (280) & \textbf{6.40 (22)} & 95.35 (328) \\

\midrule
\multirow{1}{*}{\textbf{GPT-4o}} 
& Agentless & 15.17 (44) & 98.97 (287) & 4.36 (15) & 95.35 (328) \\

\bottomrule
\end{tabular}
\end{adjustbox}
\end{table}

%% file: Figures/fig_understand.tex
\begin{figure}
    \centering
    \includegraphics[width=\linewidth]{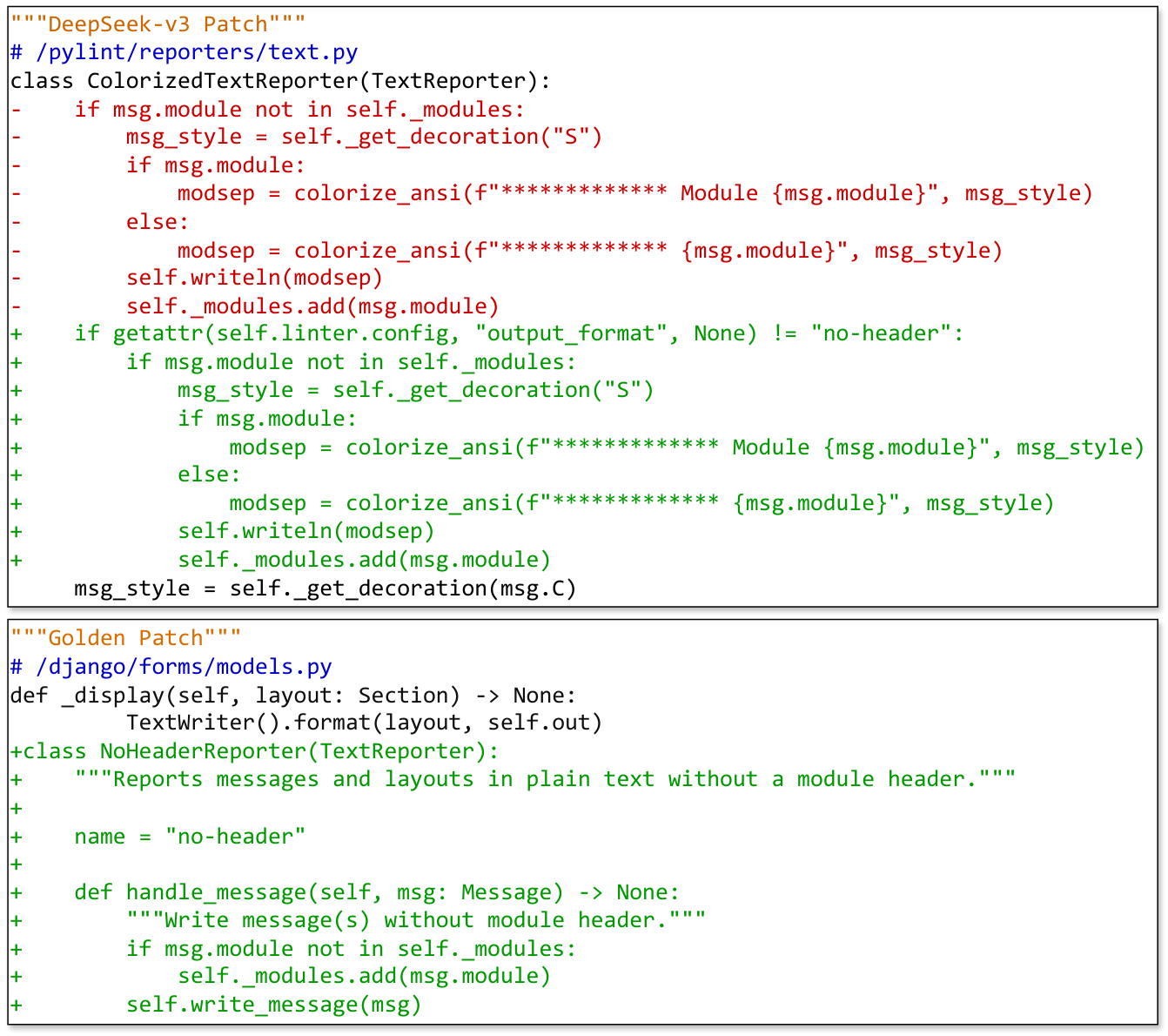}
    \caption{A failure example of \emph{pylint-dev\_\_pylint-7869} caused by an agent's lack of architectural understanding.}
    \label{fig:understand}
\end{figure}

%% file: Figures/fig_understand_doc.tex
\begin{figure}
    \centering
    \includegraphics[width=1\linewidth]{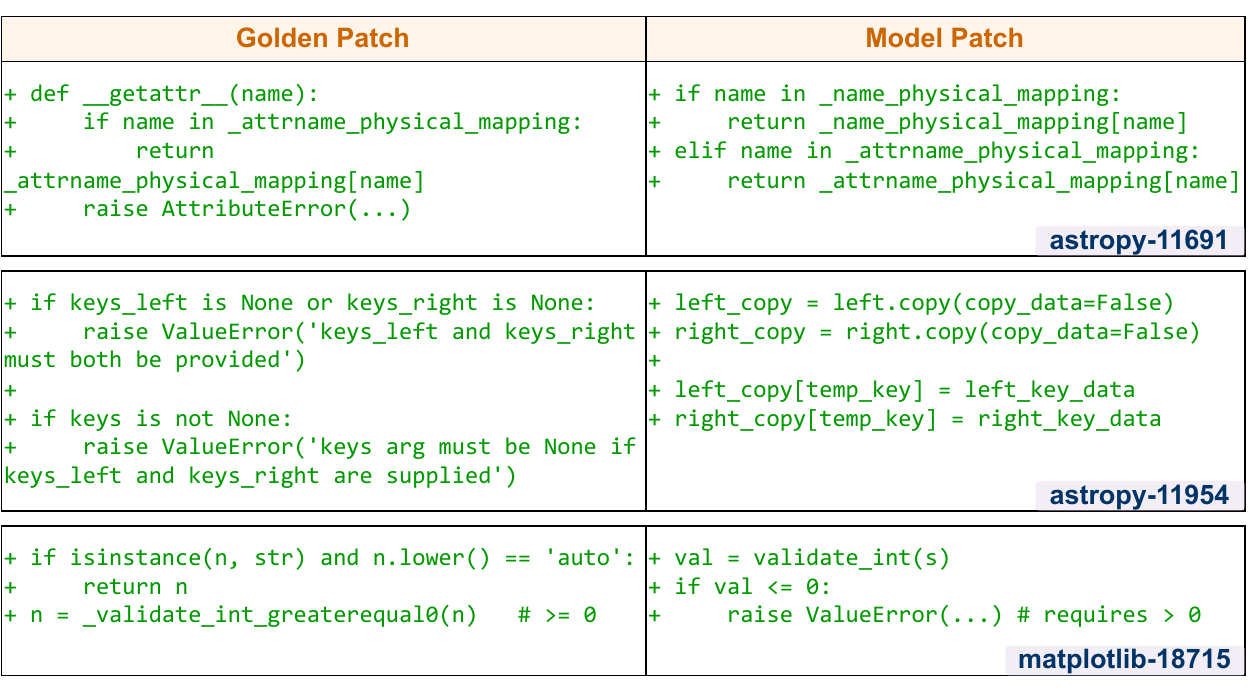}
    \caption{Representative cases of documentation semantic misinterpretation.}
    \label{fig:semantic_misunderstanding}
\end{figure}

%% file: Sections/06_discussion.tex
\section{Discussion}\label{sec:discussion}

\subsection{\rev{\nc Verified vs.\ Full}}\label{sec:full-quality}
\rev{The verification process in Section~\ref{sec:annotation-results} shows that 124 of the 238 sampled candidates (52.1\%) were filtered out, suggesting that about half of the instances in \nc Full may contain unclear task descriptions or inaccurate evaluations.
These quality issues reflect the realistic and naturalistic nature of tasks mined from real repositories, as PRs are created for human collaboration and are often iteratively clarified through maintainer-contributor discussions.
Retaining these instances in \nc Full allows it to preserve this naturalistic distribution.
Specifically, \nc Full enables evaluating agents' or models' robustness to the noise and data-quality variation of real development data, and avoids the selection bias toward simpler tasks introduced by strict filtering.
In addition, \nc Full maintains the validity of the evaluation, as model rankings remain largely consistent between the two sets (RQ2).
In summary, the two sets provide different evaluation perspectives.
\nc Verified is suited to controlled experiments, fair comparison across models, and evaluations that require high-quality annotations.
In contrast, \nc Full is suited to studying the realistic distribution of development data and robustness to noise, and thus better reflects the effectiveness of agents in real-world development.
}

\subsection{Threats to Validity}
\textbf{Internal Validity.}
\textit{Task Quality.}
In practice, the documentation change in a PR may be incomplete or not clear enough to implement the feature implemented in this PR. 
Also, the test suite of a project may not fully test a feature. To mitigate this threat, we curate a manually validated subset, \nc Verified, to increase the reliability of the evaluation.
\textit{Data Leakage.}
The instances in \nc are collected from historical commits in specific GitHub repositories. As GitHub data is widely used to train state-of-the-art models, there is a potential risk of data leakage.
Existing benchmarks that are constructed from GitHub data all face this threat.
To help mitigate this risk, we mask PR numbers during dataset construction.
The low success rates of the best-performing systems suggest the impact of data leakage is limited.
In addition, as a sanity check, we follow the methodology in \cite{ahmed2025otter} and compute Levenshtein similarity between generated patches and ground-truth patches for all successful instances of the best-performing model (DeepSeek-R1) on \nc Full. 
The average similarity is only 0.48, suggesting systems are not simply copying memorized patches.

\textbf{External Validity.}
\textit{Generalization.}
Our benchmark currently focuses on 
well-maintained open-source Python projects that provide structured release notes. While this design ensures high-quality and interpretable data, it inevitably restricts project diversity and language coverage. However, Python is one of the most popular programming languages, and the construction methodology of \nc is language-agnostic and can be adapted to other ecosystems in future work.
\textit{Scaffold Selection.}
We evaluate the performance of the selected systems using only the Agentless and OpenHands scaffolds.
Their performance on other scaffolds remains unknown. 
However, Agentless and OpenHands represent the state-of-the-art in their respective categories.
Their limited performance across different models indicates that \nc poses new challenges for agentic software engineering methods.

%% file: Sections/07_related_work.tex
\section{Related Work}
\subsection{Benchmarks for Agentic SE Systems}

Evaluating the capabilities of agentic systems in software engineering (SE) has attracted increasing attention, leading to the development of several benchmarks.
Among these, SWE-bench~\cite{jimenez2024swebench} has emerged as one of the most influential benchmarks for assessing systems on issue resolution tasks.
A line of follow-up work has extended \swe to multilingual~\cite{zan2024swe,zan2025multiswebench}, multimodal~\cite{yang2024swebenchmultimodal,guo2025omnigirl}, and industrial~\cite{rondon2025evaluating, ye2025solcontracteval} scenarios, promoting broader language and scenario coverage in issue-solving tasks.
FEA-Bench~\cite{li2025feabenchbenchmarkevaluatingrepositorylevel} aims to evaluate systems on feature addition tasks, with detailed code-level specifications, docstrings, and pull request descriptions as task inputs.
However, these benchmarks rely on maintainer-facing inputs, such as issue reports, and assume familiarity with the codebase and implementation-level terminology~\cite{chaparro2019assessing,github_issues_feature}. As a result, they provide limited insight into capabilities under public-facing specification inputs. In addition, most of them focus on bug fixing or issue solving, while largely overlooking feature addition, which accounts for about 60\% of software maintenance activities~\cite{8802807}.

Table~\ref{tab:benchmarks-comparison} provides a detailed comparison between \nc and related benchmarks, highlighting several distinctive features of \nc: 
(1) \nc adopts a documentation-driven setting where public-facing documentation changes are provided as inputs; (2) \nc does not constrain instances to those introducing new components, recognizing that many feature additions extend existing code; (3) \nc identifies feature-adding PRs based on developer-maintained release notes to reduce noise~\cite{xu2024hallucination,zhang2023siren}.
In addition, we manually construct a verified subset to enable lightweight yet reliable evaluation.

\input{Tables/table_related}

\subsection{Automating SE Tasks with Agentic Methods}

Recently, agentic methods~\cite{jiang2025agentic, jiang2025issue, wang2026icore, nashid2025issue2test} have been proposed for automating SE tasks, mainly including pipeline-based and agent-based methods.
Pipeline-based methods typically decompose tasks into stages (e.g., localization, repair, and validation) based on prior knowledge and require an underlying model to follow a fixed workflow.
Examples include Agentless~\cite{agentless}, and PatchPilot~\cite{li2025patchpilot}.
Agent-based methods equip agents with various tools (e.g., bash and language servers) to access and modify the codebase, and rely on model-driven decision-making and tool invocation.
Examples include OpenHands~\cite{wang2025openhands}, RepairAgent~\cite{icse2025-RepairAgent}, and ExecutionAgent~\cite{issta2025_ExecutionAgent}.
While these methods have achieved significant advances, they are typically evaluated in a maintainer-facing way, where the inputs include implementation-level details and assume programming expertise.
In contrast, feature addition tasks in \nc are specified with public-facing documentation changes.
This mismatch introduces new and unique challenges for applying prior methods to \nc.

%% file: Tables/table_related.tex
\begin{table}[!t]
\centering
\caption{Comparison with existing benchmarks.}
\begin{adjustbox}{width=\columnwidth}
\setlength{\tabcolsep}{2pt}
\begin{tabular}{lllll}
    \toprule
    \textbf{Aspect} & \textbf{\swe} & \textbf{GITS-Eval}   &\textbf{FEA-Bench} 
    &\textbf{\nc} 
    \\
        \midrule
    \makecell[l]{Development\\ scenario} &
    \makecell[l]{Maintainer-facing} &
    \makecell[l]{Maintainer-facing}   &\makecell[l]{Maintainer-facing} 
    &\makecell[l]{Public-facing} 
    \\
    
    \midrule
    Task type &
    \makecell[l]{Issue resolution} &
    \makecell[l]{Issue resolution}   &\makecell[l]{Feature addition} 
    &\makecell[l]{Feature addition}     
    
    \\

    \midrule
    Task input &
    \makecell[l]{Issue\\description} &
    \makecell[l]{Issue\\description}   &\makecell[l]{Code-level specs,\\ PR description} 
    &\makecell[l]{Documentation\\changes} 
    \\
    
    \midrule
    Instance selection & N/A & N/A & \makecell[l]{PRs introducing\\new components} 
    & \makecell[l]{PRs containing\\doc changes}
    \\
    \midrule
    \makecell[l]{Feature\\ identification} & N/A & N/A &\makecell[l]{Prompting\\ models} 
    &\makecell[l]{Mining\\release notes} 
    \\
    \midrule
    Verified subset & Yes & Yes   & No 
    &Yes 
    \\
    \midrule
    Open-source & Yes & No   &Yes &Yes \\
    \bottomrule
\end{tabular}
\end{adjustbox}
\label{tab:benchmarks-comparison}
\end{table}

%% file: Sections/08_conclusion.tex
\section{CONCLUSION}

We introduce \nc, a challenging benchmark for evaluating software engineering capabilities in documentation-driven feature addition.
\nc is constructed with a systematic five-phase pipeline.
It consists of 634 real-world feature addition tasks identified from developer-maintained release notes.
Given documentation changes within a repository as input, \nc requires systems to generate a patch that adds a new feature.
We evaluate 8 advanced models using two representative scaffolds, Agentless and OpenHands.
Our results show that documentation-driven feature addition is a uniquely difficult and unsolved problem.
Moreover, we analyze failure cases on \nc to guide future improvements in cross-file editing, code module understanding, and documentation semantic interpretation.